\documentclass[preprint2]{aastex}
\usepackage{graphicx}
\usepackage{picture,calc} 
\usepackage{amssymb,amsmath,pifont}
\usepackage{url}
\usepackage{hyperref}

\usepackage{color}

\def\be{\begin{equation}}
\def\ee{\end{equation}}

\def\ba{\begin{align}}
\def\ea{\end{align}}

\shorttitle{The Synchronicity Problem}
\shortauthors{Avelino et al.}

\begin{document}

\title{The dimensionless age of the Universe: a riddle for our time}
\author{Arturo Avelino\altaffilmark{1} and Robert P. Kirshner\altaffilmark{1,2}}
\altaffiltext{1}{Harvard-Smithsonian Center for Astrophysics, 60 Garden Street, Cambridge, Massachusetts, 02138, USA}
\altaffiltext{2}{Gordon and Betty Moore Foundation, 1661 Page Mill Road, Palo Alto, CA 94304}
\email{aavelino@cfa.harvard.edu}

\begin{abstract}
{ We present the interesting coincidence of cosmology and astrophysics that points toward a dimensionless age of the universe $H_0t_0$ that is close to one.  Despite cosmic deceleration for 9 Gyr and acceleration since then, we find $H_0t_0 = 0.96 \pm 0.01$ for the $\Lambda$CDM model that fits SN Ia data from Pan-STARRS, CMB power spectra, and baryon acoustic oscillations.  Similarly, astrophysical measures of stellar ages and the Hubble constant derived from redshifts and distances point to $H_0 t \sim 1.0 \pm 0.1$.  The wide range of possible values for $H_0t_0$ realized during comic evolution means that we live at what appears to be a special time.  This ``synchronicity problem'' is not precisely the same as the usual \textit{coincidence problem}  because there are combinations of $\Omega_M$ and $\Omega_{\Lambda}$ for which the usual coincidence problem holds but for which $H_0t_0$ is not  close to 1. }
\end{abstract}



		\section{Introduction}

In the years just before the discovery of cosmic acceleration, it was hard to reconcile the age of the Universe  with the ages of stars.  The cosmic expansion time implied by a Hubble Constant $H_0$ of about 70 km s$^{-1}$ Mpc$^{-1}$ or more \citep{RiessPressKirshner1996,FreedmanEtal1994,SchmidtEtal1994} and the theoretically favored spatially flat cosmological model with $\Omega_M =1$,  was 9 Gyr.  The ages of the oldest stars were estimated to be greater than 12 Gyr \citep{ChaboyerEtal1996a,ChaboyerEtal1996b}.  You should not be older than your mother, and the objects in the Universe should not be older than the time from the Big Bang.
The discovery of cosmic acceleration solved this riddle: with a cosmological constant of the amount required by the supernova observations, $\Omega_{\Lambda} \sim 0.7$ \citep{Riess-DiscoveryAccelUniverse:1998,Perlmutter-DiscoveryAccelUniverse1998}, 
the age of the Universe implied by a Hubble constant near 70 km s$^{-1}$Mpc$^{-1}$ was about 14 Gyr, in good accord with the ages inferred from stellar evolution in globular clusters \citep{AgeUniv-CarretaEtal-2000}.

But the new results posed their own conundrum.  Despite a Universe that was decelerating for its first 9 Gyr, then shifting through cosmic jerk to acceleration for the past 5 Gyr, the present value of the dimensionless age, $H_0t_0$ was constrained by supernova distances to lie eerily close to 1 (for instance, to $H_0t_0=0.96 \pm  0.04$ as estimated by  \citet{TornyEtal2003}).   This is strange because, as we show in section 2 of this paper, over the span of cosmic time, the dimensionless age of the Universe can take on a wide range of values.   It appears that we are living today at a privileged time when the dimensionless age, at least for a $\Lambda$CDM Universe, is very close to one.

This problem of the dimensionless age is similar to, but not identical with the well-known puzzle of the \textit{coincidence problem}, in which $\Omega_M$ and $\Omega_{\Lambda}$ are nearly equal now, though they were not in the past and will not be in the future of a $\Lambda$CDM universe.   It is different because, as we show below, it is possible to have combinations of $\Omega_M$ and $\Omega_{\Lambda}$ for which the coincidence problem persists but where $H_0t_0$ departs from 1.

Another curious feature of the $\Lambda$CDM model is the possibility that the cosmological constant might actually be the quantum vacuum energy of gravity \citep{WeinbergCosmoConstProblem,CarrollPressTurner1992-CosmologicalConstant}.  This idea is attractive because the equation of state of the cosmological constant corresponds to a fluid with negative pressure in just the same way as the equation of state for the vacuum energy in particle physics.  However, there is a large discrepancy $(10^{120})$ between the value of the energy density from cosmological observations and the value of the vacuum energy from quantum theory \citep{Carroll-2001-CosmologicalConstant}.  This is the cosmological constant problem for which solutions have been sought with anthropic arguments (see for instance \citet{Weinberg2000} and references therein) and by appeal to a vast landscape of possible universes in a multiverse (for instance \citet{Kragh2009-Multiverse}).

The physical nature of the cosmological constant remains obscure.  Einstein introduced the cosmological constant at a time when common wisdom held that the Universe was the Milky Way.  He said, ``\textit{That term is necessary only for the purpose of making a quasi-static distribution of matter, as required by the fact of the small velocities of the stars.}'' \citep{Einstein1917}.  
 The subsequent discovery of the large universe of galaxies and its cosmic expansion led to his decision, with de Sitter, to leave the cosmological constant out of future cosmological work \citep{Einstein-deSitter-1932}.

Because there is no underlying physical theory, the $\Lambda$CDM model does not provide any information on the nature of $\Lambda$.   Even if future observations constrain dark energy to have the equation-of-state of the cosmological constant $(w = -1)$ to arbitrary precision, we will still need to understand it.

Section \ref{SectionLCDM} of this paper defines the dimensionless age of the Universe and works out its variation over time in $\Lambda$CDM cosmologies.  Section \ref{SectionLCDMObservations} presents the evidence that the current best value for the dimensionless age is close to 1.  Section \ref{SectionAstrophysics} compares astrophysical constraints on the age of objects in the universe with results inferred from the expansion history.  Section \ref{SectionConclusions} presents our conclusions.

	\section{The time coincidence problem in $\Lambda$CDM model}
	\label{SectionLCDM}

The age of the Universe, $t$, is the time elapsed since the Big Bang when the scale factor was $a=0$.  
Its expression in terms of the scale factor can be easily found from the definition of the Hubble parameter $H(a) \equiv \dot{a}/a$:
\begin{equation}\label{EqIntegraDef}
H(a) = \frac{1}{a}\frac{da}{dt}   \, \, \, \Rightarrow \, \, \, t(a) = \int^a_0 \frac{da'}{a' H(a')},
\end{equation}
where we have assumed $t(a=0) =0$. To calculate the \textit{present} age of the Universe, $t_0$, we have to compute the integral (\ref{EqIntegraDef}) in the interval\footnote{For $a=0$ the Eq. (\ref{EqIntegraDef}) is singular, so the lower limit of the integration is $a>0$.} $a=(0, 1]$.

For the base $\Lambda$CDM model,  we have that the Hubble parameter has the form 
\be \label{EqHubbleParameter}
H^2(a) = H^2_0 \left( \frac{\Omega_{\rm M}}{a^3} + \Omega_{\Lambda} + \frac{\Omega_{\rm r}}{a^4} + \frac{\Omega_{\rm k}}{a^2} \right),
\ee
where $\Omega_{\rm M}$, $\Omega_{\Lambda}$,  $\Omega_{\rm r}$ and $\Omega_{\rm k}$  are the matter (baryon and cold dark matter), cosmological constant, radiation and curvature components. 
Using Eq. (\ref{EqHubbleParameter}) we can express  Eq. (\ref{EqIntegraDef}) as 
\begin{equation}\label{EqIntegralForHoto} 
H_0 t(a) = \int^a_0 \frac{da'}{a' \sqrt{\Omega_{\rm M}/a'^3 + \Omega_{\Lambda} + \Omega_{\rm r}/a'^4 + \Omega_{\rm k}/a'^2}},
\end{equation}
where $H_0 t(a)$  is the \textit{dimensionless age} of the Universe. 

From  Eq. (\ref{EqIntegralForHoto}),  if the density of all the components of the Universe are known, then $H_0 t_0$ is known too. Conversely, if $t_0$ and $H_0$ are measured independently, then we obtain a measurement of the combination of the cosmological parameters.

For the simple case of a spatially-flat Universe composed of matter alone at late times, equation (\ref{EqIntegralForHoto}) would be
\be\label{EqHoto23Omega}
H_0t(a) = \frac{2 a^{3/2}}{3 \sqrt{\Omega_M}}.
\ee
From Eq. (\ref{EqHoto23Omega}), for the values of $\Omega_M =1$, $a=1$ and $H_0$ = 72 km s$^{-1}$Mpc$^{-1}$, the present age of the Universe would be $t_0 = 9.05$ Gyr.

On the other hand, the very early Universe was dominated by radiation. At that time the Hubble parameter can be simply expressed as  $H(a) = H_0 \sqrt{\Omega_r /a^4}$, and then $H_0t(a)$ given in equation (\ref{EqIntegralForHoto}) becomes 
\begin{align}
H_0 t(a \rightarrow 0) &=  \frac{1}{\sqrt{ \Omega_{r}}} \int^{a \rightarrow 0}_0 a' da', \\
\Rightarrow \quad  H_0 t(a \rightarrow 0) & \rightarrow 0.
\end{align}

Also, as the Universe evolves to a dark energy-dominated epoch  ($a \rightarrow \infty$),  the  Hubble parameter  becomes  $H(a) = H_0 \sqrt{\Omega_{\Lambda}}$, and then  equation (\ref{EqIntegralForHoto}) yields
\begin{align}
H_0 t(a \rightarrow \infty) &=  \frac{1}{\sqrt{ \Omega_{\Lambda}}} \int^{\infty}_0 \frac{da'}{a'}, \\
\Rightarrow \quad H_0 t(a \rightarrow \infty) &  \rightarrow \infty.
\end{align}

So, during the  evolution of a physically reasonable universe, the range of dimensionless ages corresponds to $0 < H_0t(a) < \infty$.

At the present time ($a = 1$) the value of  $H_0t_0 \equiv H_0 t(a=1) $ is fixed by the present matter-energy content in the Universe, i.e., for  different values of ($\Omega_{\rm M}, \Omega_{\Lambda}, \Omega_{\rm r}, \Omega_{\rm k}$) we obtain different histories for the evolution of $H_0 t(a)$, and different {present values} for $H_0 t_0$.

Given that the current CMB data constrain $\Omega_k$ to be near zero and that $\Omega_{\rm r}  \ll 1$ today,  a good approximation for Eq. (\ref{EqIntegralForHoto}) evaluated at the present time is
\begin{equation}\label{EqIntegralForHotoFlat} 
H_0 t_0  = \int^1_0 \frac{da}{a \sqrt{\Omega_{\rm M}/a^3 + (1- \Omega_{\rm M})}},
\end{equation}
where we can see that $H_0 t_0 = H_0 t_0(\Omega_M)$. When $\Omega_M \rightarrow 0$ then
\be \label{EqHotoLimitInfinity}
H_0 t_0(\Omega_M \rightarrow 0) = \int^1_0 \frac{da}{a} = \infty.
\ee
And when $\Omega_M \rightarrow 1$ then
\be  \label{EqHotoLimitZero}
H_0 t_0(\Omega_M \rightarrow 1) = \int^1_0 \sqrt{a} \, \, da = \frac{2}{3}.
\ee
So, the \textit{present}  dimensionless age of the Universe can have values in the range $2/3 < H_0t_0 < \infty$, depending on the magnitudes of the cosmological parameters.

Despite this very large range for the value of $H_0 t_0$ and  the very large range for $H_0 t(a)$ along the whole evolution of the Universe ($0<H_0 t(a) < \infty$), it turns out that  $H_0 t_0 \sim 1$, according to the $\Lambda$CDM model combined with the current cosmological observations.

\begin{figure}
\begin{center}
\includegraphics[width=7cm]{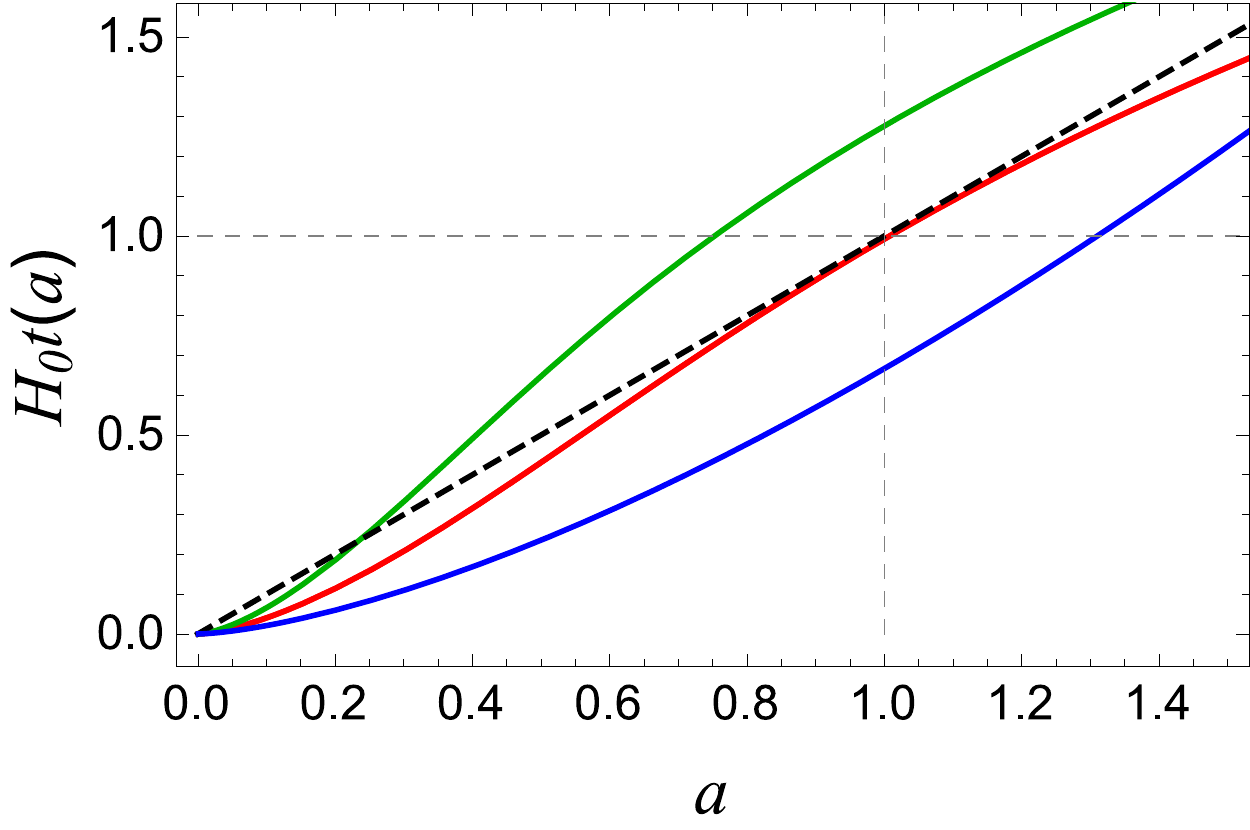}
\caption{(color online)   Evolution of $H_0t(a)$ as a function of the scale factor $a$. The red  line corresponds to the values of $(\Omega_{\rm M} = 0.27, \Omega_{\Lambda} = 0.73,  w = -1)$, while the { blue and green solid lines to the arbitrary values of $(1, 0, -1)$ and $(0.1, 0.9, -1)$ respectively to illustrate the differences produced by the cosmological parameters. 
The black line corresponds to a constant value of $\dot{a}$.  In that case the age of the Universe is always equal to the Hubble time:  $H_0\, t = 1$. }
}  
\label{FigureEvolutions}
\end{center}
\end{figure}

\begin{figure}
\begin{center}
\includegraphics[width=7cm]{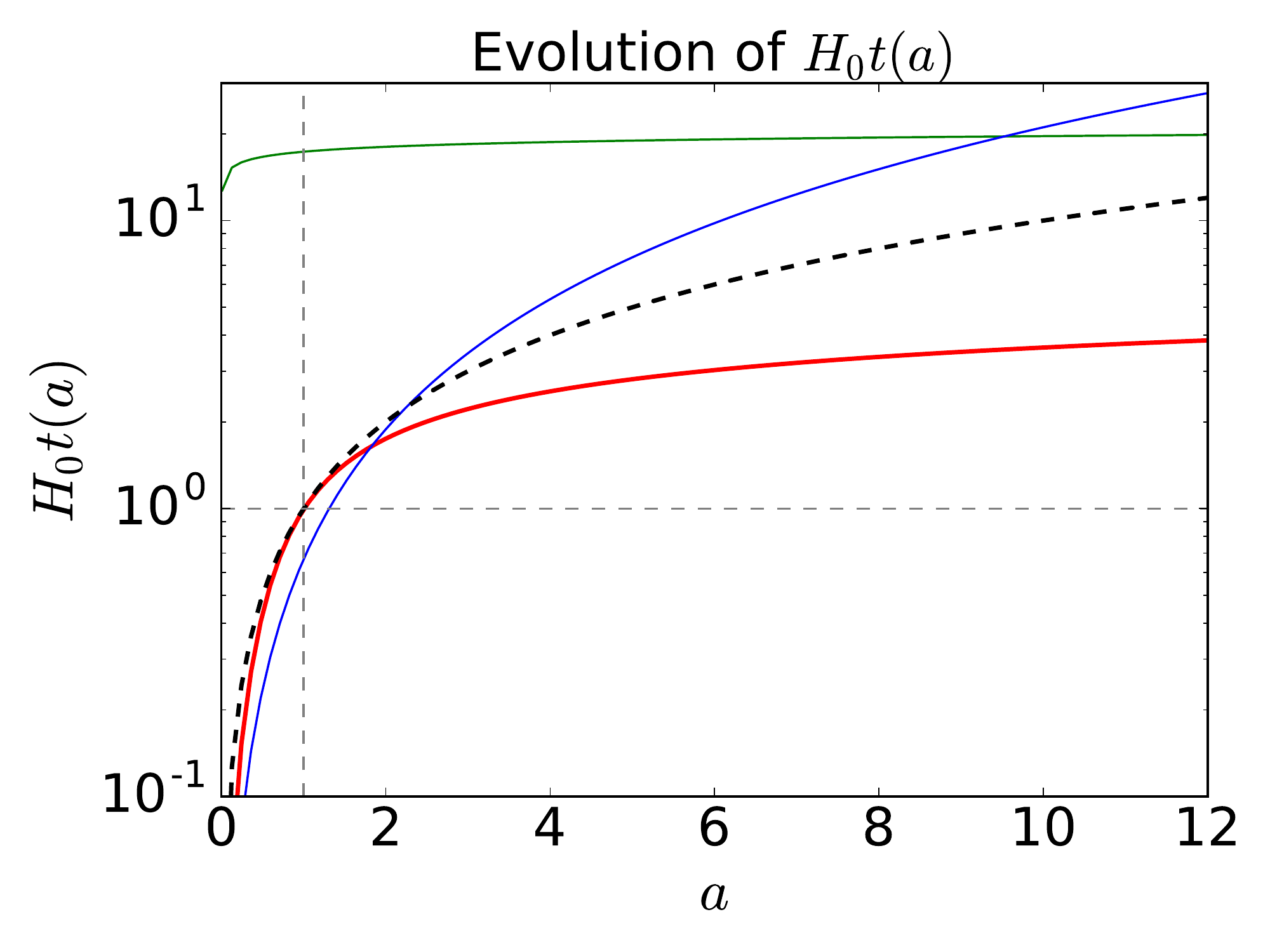}
\caption{(color online)   Evolution of $H_0t(a)$ as a function of the scale factor $a$ in log scale. 
In this plot the green line corresponds to the arbitrary values of  ($\approx$0$, 1,-1)$. 
}
\label{FigureEvolutions2}
\end{center}
\end{figure}

Figures \ref{FigureEvolutions} and  \ref{FigureEvolutions2} show the evolution of $H_0 t(a)$ defined in equation (\ref{EqIntegralForHoto}).
We can read these figures as follows: at the fixed value $a=1$ (vertical dashed line), from a theoretical point of view (i.e., before considering the cosmological observations), and for any given physical values of the cosmological parameters of $\Lambda$CDM  in a wide range, 
the present value of  $H_0t_0 = H_0t(a=1) $ can have a broad range of positive values located at any part on the line $a=1$ in the interval  $2/3<H_0t_0<\infty$. 
However, when the observations are used to constrain the cosmological parameters in $\Lambda$CDM, it happens that the \textit{present} value of the dimensionless age is very close to $H_0t_0 =1.0$ (see the intersection of the red curve with the line $a=1$ in Fig. \ref{FigureEvolutions}).

For a wide range of values of the cosmological parameters, there is some value $a_*$ where $H_0t_0$ is 1. In Fig. \ref{FigureEvolutions}, that is the place where the black solid line crosses the green band. The actual cosmological parameters combined with $\Lambda$CDM indicates that time is today, when $a_* =1$. We call this coincidence ``the synchronicity problem''.

{ The identity  $H_0t_0 = 1$ is  equivalent to saying  that the  age of the Universe is exactly equal to the \textit{Hubble time}, that corresponds to a Universe expanding at the same rate always since the Big Bang, i.e., $\dot{a} = $ constant. Given that $H(a) \equiv \dot{a}/a$ then $H(a=1) =  \text{constant}/1 = H_0$, thus
\be \label{EqScaleFactorRate1}
\dot{a} = H_0
\ee
Thus, if the Universe were expanding at the constant rate $H_0$ since its beginning, then its age would be $1/H_0$. }

{In this scenario, the dimensionless age evolves as
\be \label{EqHoHubbleTime}
H_0 t(a) = a
\ee
Figures \ref{FigureEvolutions} and \ref{FigureEvolutions2} show this scenario with a black dashed line.}

{ So, the fact that the dimensionless age is close to one today implies that the early decelerated expansion epoch of the Universe is  compensated by the late time accelerated phase so that the total expansion \textit{today} is almost exactly as if the Universe were expanding at the \textit{same} expansion rate $\dot{a} = $ constant since the Big Bang. Is this a ``coincidence''?}

{This coincidence only happens today: for any other time in the past or future the age of the Universe at that time will not be so similar.  
We discuss  these ideas further and compare with astrophysical observations at the end of section \ref{SectionAstrophysics}. }

{ Figures \ref{FigureEvolutions} and \ref{FigureEvolutions2} shows how the red line (the actual expansion evolution of the Universe) and black dashed line (a Universe expanding at constant rate) intercept or are very close each other \textit{today} compared with the past or future evolution of the Universe, i.e., the actual age of the Universe is very close to the Hubble time. }

On the other hand,  a known cosmological model that predicts a present value of the dimensionless age equal to
one corresponds to the  \textit{Milne  model} \citep{MilneUniverse-1935}, composed of an empty Universe (zero matter-energy density) with a spatially negative curvature $\Omega_{\rm k} =1$.
From Eq. (\ref{EqIntegralForHoto}), we can see that for this model the present dimensionless age corresponds to
\be
H_0t_0 = \int^1_0 \frac{da'}{a' \sqrt{\Omega_{\rm k}/a'^2}} = 1.
\ee
However, the Milne model is clearly ruled out by the current cosmological observations.

We can also visualize this `time coincidence' by plotting the integrand of Eq. (\ref{EqIntegralForHoto}),  %
\begin{equation}\label{EqIntegrandForHoto} 
F(a) \equiv  \frac{1}{a \sqrt{\Omega_{\rm M}/a^3 + \Omega_{\Lambda} + \Omega_{\rm r}/a^4 + \Omega_{\rm k}/a^2}}.
\end{equation}
{ For the case of a Universe expanding always at the same rate since the Big Bang then Eq. (\ref{EqIntegrandForHoto}) simply becomes $F(a) = 1$.}

\begin{figure}
\begin{center}
\includegraphics[width=7cm]{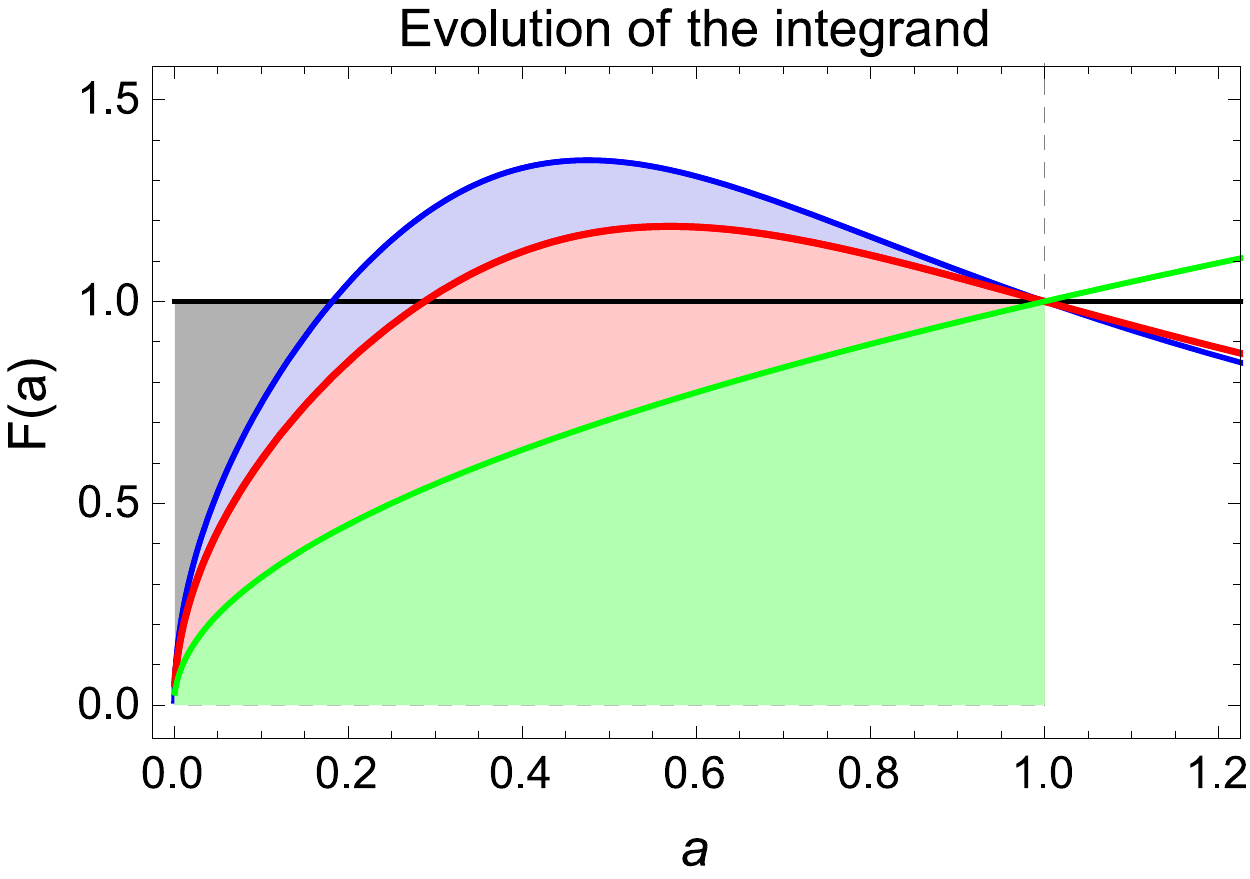}
\caption{(color online)  Evolution of the function $F(a)$ defined in equation (\ref{EqIntegrandForHoto}), that corresponds to the integrand  of the Eq. (\ref{EqIntegralForHoto}). 
The red  line corresponds to the values of $(\Omega_{\rm M} = 0.27, \Omega_{\Lambda} = 0.73,  w = -1)$, while the blue and green lines correspond to the arbitrary values of $(0.18, 0.84,-1)$, $(0.4, 0.6, -1)$ respectively, to illustrate the differences between some models. 
The area under these curves in the interval $0<a<1$ gives the value of $H_0t_0$. 
For the red curve the area  is equal to 1.
Notice that for different values of the cosmological parameters, the area under  $F(a)$ can be very different.
{ The black line corresponds to $F(a) = 1$, a universe that expands with constant $\dot{a}$.  The Synchronicity problem depicted graphically is that the gray area and the red are very nearly equal.}
 }
\label{FigureEvolutionsIntegrand}
\end{center}
\end{figure}

In Fig. \ref{FigureEvolutionsIntegrand}, the fact that $H_0 t_0 \sim 1$ means that the area under the curve of $F(a)$ in the interval $0<a<1$ is  { about 1 (area below the red curve and black horizontal line),} despite the fact that this area might have any other value (as shown by the other colored curves), before considering the observations.

	\section{Observational evidence of $H_0t_0 \approx 1$ in $\Lambda$CDM}
	\label{SectionLCDMObservations}	

To illustrate how $H_0 t_0 \sim 1$ according to the cosmological observations 
we use the type Ia supernova (SN) sample of PanSTARRS \citep{SNe-Panstarrs1-Rest_Kirshner2013} (hereinafter PS1),  combined with the baryon acoustic oscillations (BAO) and the cosmic microwave background radiation (CMB) data, to constrain the free parameters of the $\Lambda$CDM model and then compute the probability density function (PDF) of $H_0t_0$.

We use also the values for the cosmological parameters reported by 
\citet{CMB_Planck_Cosmology_2015} that combines the CMB data with BAO and the Joint Light-curve  Analysis (JLA) SN Ia compilation by \citet{SNe-BetouleEtal-2014}. 
The MCMC Markov chains reported by \citet{CMB_Planck_Cosmology_2015} that we used in this work are publicly available at 
\href{http://pla.esac.esa.int/pla/\#cosmology}{http://pla.esac.esa.int/pla/\#cosmology}

Table \ref{TableCosmoValues} shows the best estimated values for $(\Omega_{\rm M}, \Omega_{\Lambda})$, using these different SN compilations and implementations of the BAO and CMB data.

\begin{figure}
\begin{center}
\includegraphics[width=7.5cm]{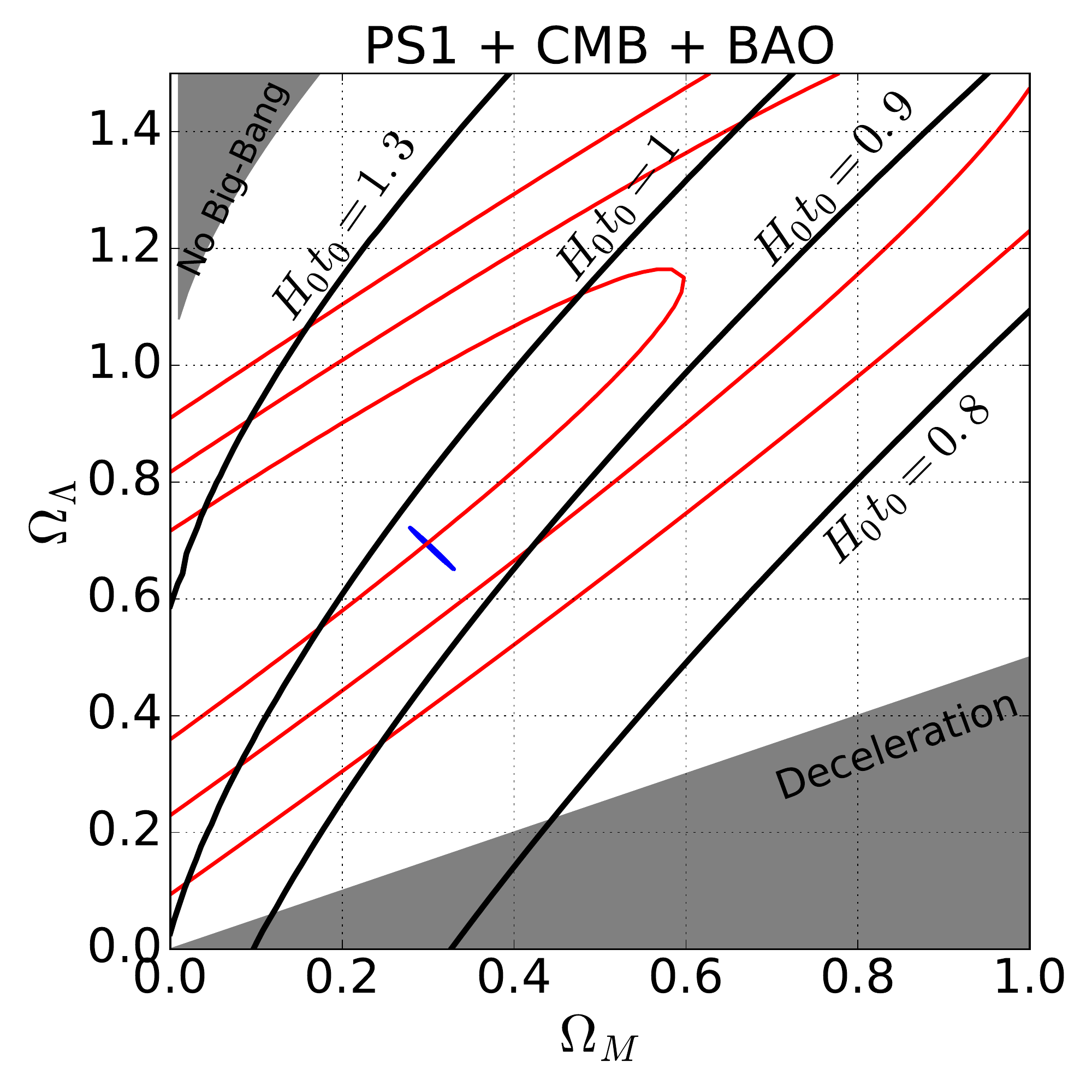}
\caption{(color online) The 1$\sigma$ and 2$\sigma$ constraints on ($\Omega_M, \Omega_{\Lambda}$) using the Pan-STARRS SN (red), CMB (dashed blue) and BAO (solid blue) data sets. The black small contours correspond to the constraints from the combined PS1+CMB+BAO data. 
The marginalized best estimates are given in the first rows of table \ref{TableCosmoValues}.
The black lines show different values for $H_0t_0$.
We notice that the contours of constraints from supernovae alone select values of $H_0t_0$ more stringently than either the BAO or the CMB constraints taken alone.
 }
\label{FigureOmOL}
\end{center}
\end{figure}

\begin{figure}
\begin{center}
\includegraphics[width=7.5cm]{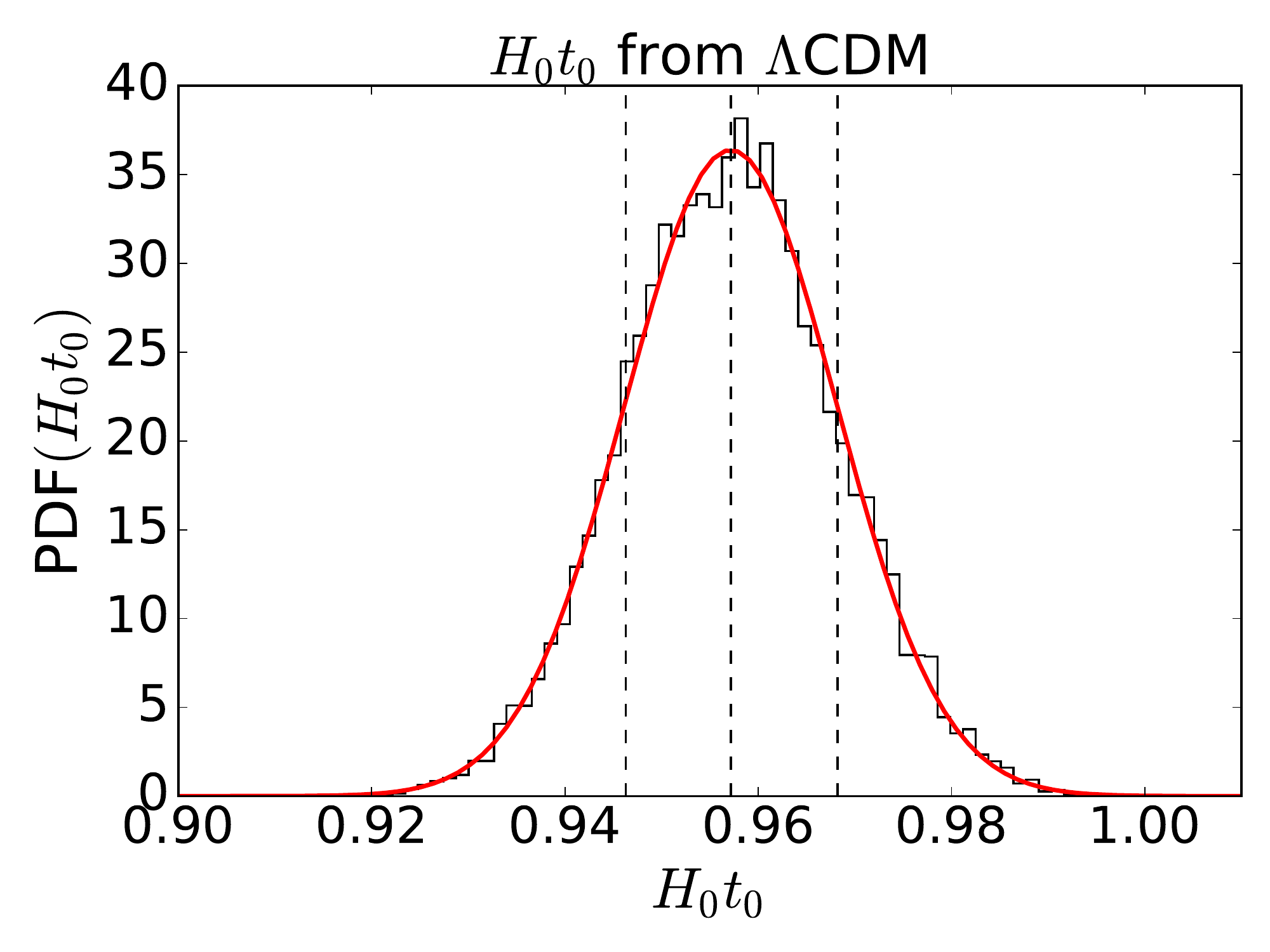}
\caption{ Estimation of the probability density function (PDF) of $H_0t_0$ (solid red curve) computed from the Markov chains of the MCMC sampling for $\Lambda$CDM using PS1+CMB+BAO data  and  equation (\ref{EqIntegralForHoto}).
The vertical dashed lines correspond to 16\%, 50\% and 84\% percentiles. 
We can see that the best estimated value of $H_0t_0$ is around 1. See table \ref{TableCosmoValues} and appendix \ref{Sec_MCMCMarkovChains}. }
\label{FigureOmOLPDF}
\end{center}
\end{figure}

Fig. \ref{FigureOmOL} shows the constraints on $(\Omega_{\rm M}, \Omega_{\Lambda})$ parameter space
from the combined PS1, BAO and CMB data sets (blue narrow contours).  
The black lines correspond to different values of $H_0t_0$ computed using  equation (\ref{EqIntegralForHoto}). 
We can see how the confidence regions from the joint SN+BAO+CMB data are close to the  $H_0t_0 = 1$ line.
We notice also that the contours of constraints from supernovae alone select values of $H_0t_0$ more stringently than either the BAO or the CMB constraints taken alone.

Fig. \ref{FigureOmOLPDF} shows the PDF of $H_0t_0$ computed from the Markov chain Monte Carlo (MCMC) sampling of the $\Lambda$CDM model and equation (\ref{EqIntegralForHoto}) (see appendix \ref{Sec_MCMCMarkovChains} for details.)
The red  curve corresponds to a Gaussian PDF with mean = 0.96 and standard deviation = 0.01 (see Table \ref{TableCosmoValues}). The vertical dashed lines correspond to 16\%, 50\% (the median) and 84\% percentiles.

\begin{table*}
\centering
\begin{tabular}{ c c c c c}
\multicolumn{5}{c}{ \textbf{Best estimated parameters and $H_0t_0$ from $\Lambda$CDM and extensions} } \\
\hline \hline
\multicolumn{5}{c}{Pan-STARRS1+CMB+BAO} \\
\hline
Label  & $\Omega_{\rm M}$ &  $\Omega_{\Lambda}$  & $w$ &  $H_0 t_0$   \\  \hline

$\Lambda$CDM & $0.30^{+0.011}_{-0.010}$ & $0.69^{+0.015}_{-0.016}$  & $-1^*$  & $0.957 \pm 0.011$    \\

flat $\Lambda$CDM &  $0.28^{+0.008}_{-0.008}$  &$^a0.72^{+0.008}_{-0.008}$ &  $-1^*$ & $0.954 \pm 0.007$  \\

$w$-CDM &  $0.28^{+0.017}_{-0.016}$ & $^a0.72^{+0.017}_{-0.016}$  & $-1.13^{+0.071}_{-0.074}$ & $1.001 \pm 0.027$  \\

\hline
 \\
\multicolumn{5}{c}{JLA + CMB + BAO} \\
\hline
$\Lambda$CDM  & $0.31 \pm 0.006$ &  $0.69 \pm 0.006 $ & $-1^*$ & $ 0.955 \pm 0.005$    \\
flat $\Lambda$CDM  & $0.312 \pm 0.009$  & $^a0.688 \pm 0.009$  & $-1^*$ &  $0.953 \pm 0.007$   \\
 $w$-CDM &  $0.306 \pm 0.009$ &  $^a0.694 \pm 0.009$  & $-1.029 \pm 0.04$ & $0.963 \pm 0.013$  \\
\hline
\end{tabular}
\caption{Best estimated values of the cosmological parameters from the SN + CMB + BAO and the inferred best estimate for $H_0t_0$. The best estimated values from JLA+CMB+BAO  were computed from the MCMC chains reported by \citet{CMB_Planck_Cosmology_2015} (see appendix). 
All the uncertainties in the table correspond to the 68.3\% confidence intervals.
* Fixed values. $^a$ Implied value: $\Omega_{\Lambda} = 1- \Omega_{\rm M} - \Omega_{\rm k} - \Omega_{\rm r}$. }
\label{TableCosmoValues}
\end{table*}


	\subsection*{Extension of $\Lambda$CDM}
	\label{SectionExtensionsToLCDM}

Now we investigate the prediction for the value of $H_0t_0$ given by the one-parameter extension to the  $\Lambda$CDM model: allowing for an arbitrary value of the constant equation-of-state parameter (EoS) $w$ for the dark energy component, in a spatially flat Universe, and labeled as $w$-CDM.

To compute this case we use the following expression 
\begin{equation}\label{EqIntegralForHotoCPL} 
H_0 t_0 = \int^1_0 \frac{da'}{a' \sqrt{\Omega_{\rm M}/a'^3 + \Omega_{\rm X}/a^{3(1+w)} + \Omega_{\rm r}/a'^4 }}, 
\end{equation}
where $\Omega_{\rm X}$ is the dark energy component. 
As usual, we have also the Friedman constraint given as $\Omega_{\rm M} +  \Omega_{\rm X} + \Omega_{\rm r} =1$. 

Table \ref{TableCosmoValues}  shows the best estimated values for $(\Omega_{\rm M}, \Omega_\Lambda,  w)$ and the implied value for $H_0t_0$ from the Eq. (\ref{EqIntegralForHotoCPL}), using the different combinations of  SN+CMB+BAO data.

\begin{figure}
\begin{center}
\includegraphics[width=7.5cm]{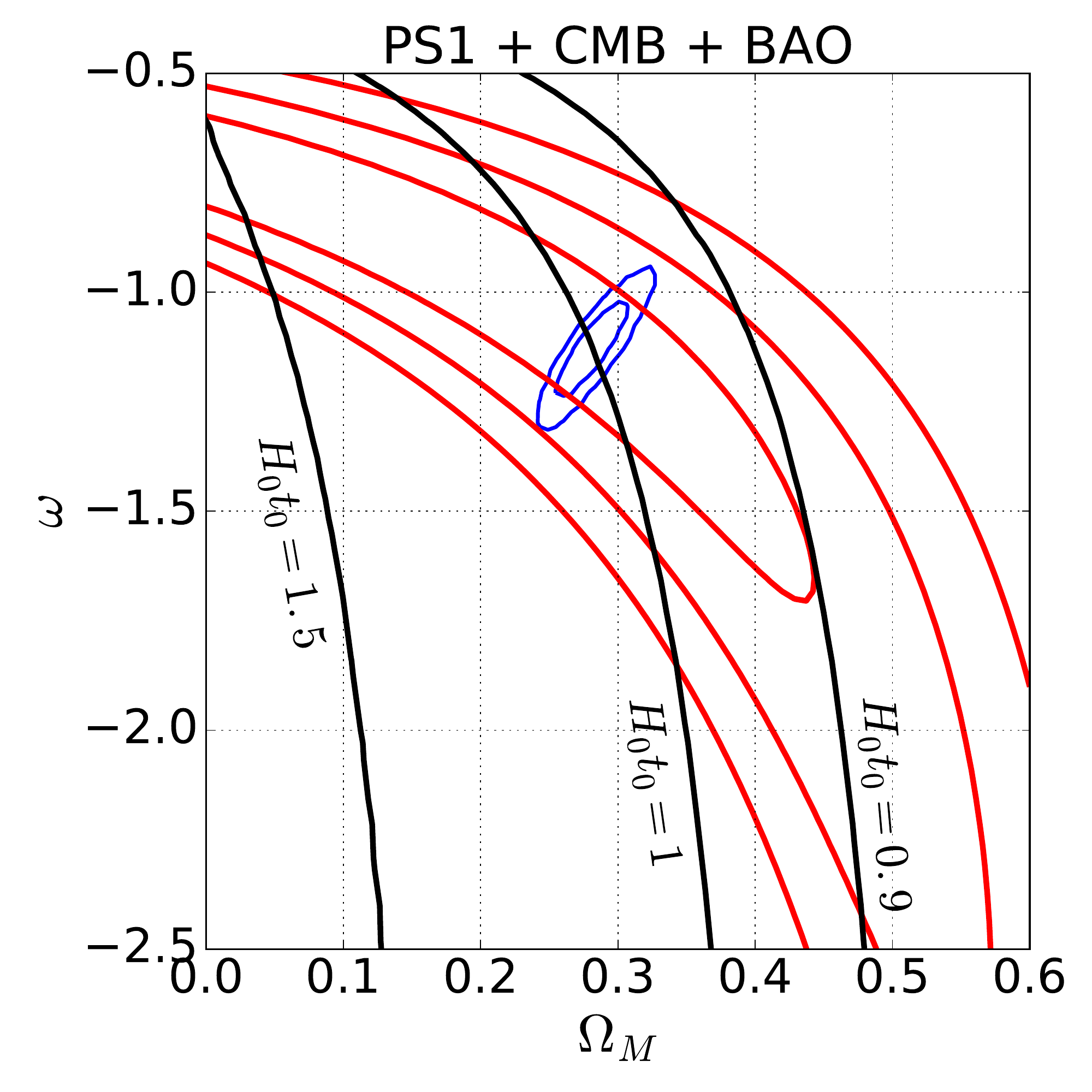}
\caption{The 1$\sigma$ and 2$\sigma$ constraints on ($\Omega_M, w$) using the PS1 + CMB + BAO data.  See table \ref{TableCosmoValues}.
The black lines show different values for $H_0t_0$. }
\label{FigureOmWde}
\end{center}
\end{figure}

\begin{figure}
\begin{center}
\includegraphics[width=7.5cm]{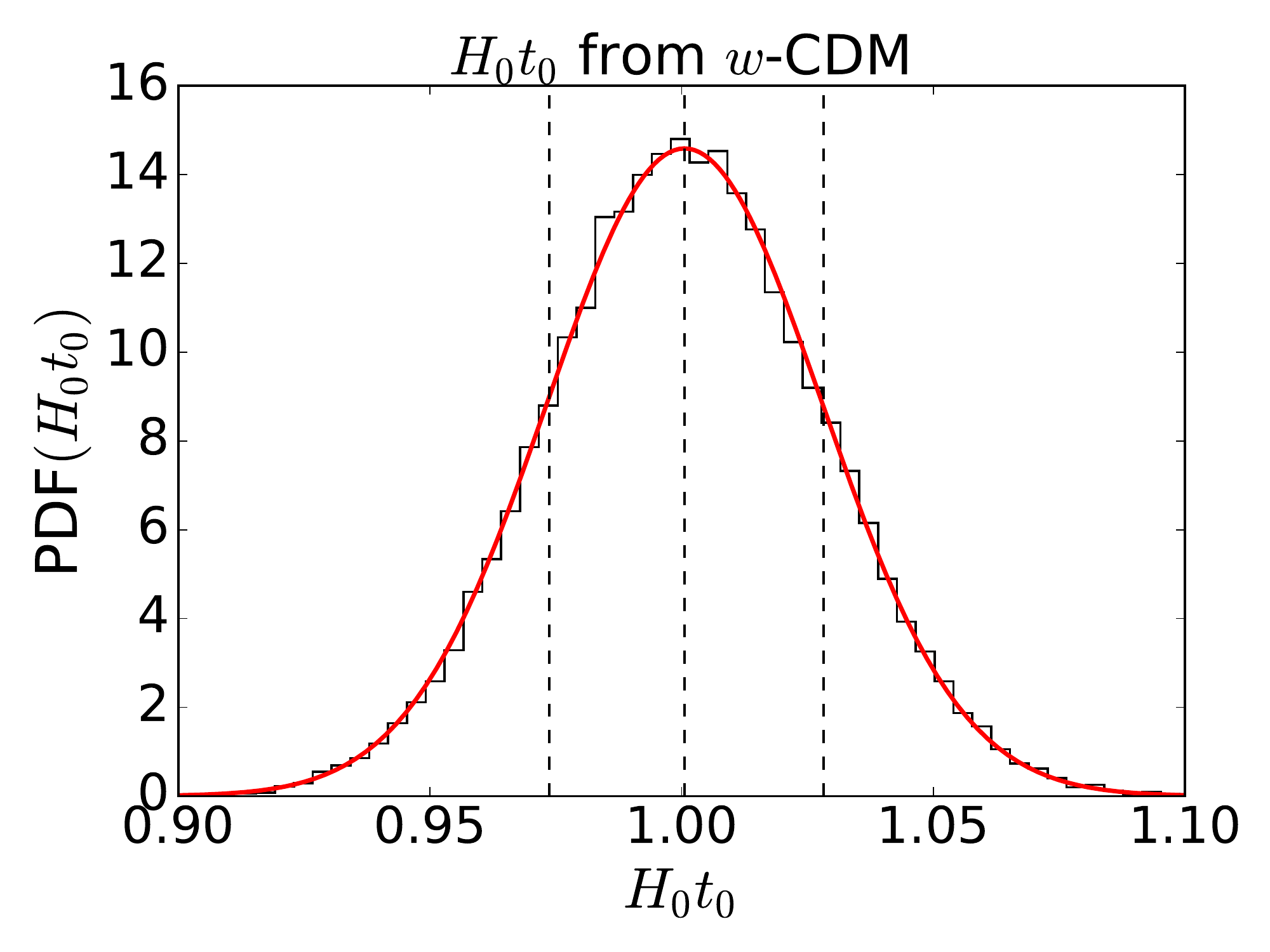}
\caption{Estimation of the PDF of $H_0t_0$ computed from the Markov chains of the MCMC sampling for $w$-CDM using PS1+CMB+BAO data  and  Eq. (\ref{EqIntegralForHotoCPL}).
The vertical dashed lines correspond to 16\%, 50\%  and 84\% percentiles. 
We observe again that the best estimated value of $H_0t_0$ is around 1.}
\label{FigureOmWdePDF}
\end{center}
\end{figure}

Fig. \ref{FigureOmWde}  shows  the constraints on the $w$-CDM model from PS1+CMB+BAO data (blue contours). 
 And Fig.  \ref{FigureOmWdePDF} shows the estimated PDF of $H_0t_0$ computed from Eq. (\ref{EqIntegralForHotoCPL}) and the MCMC Markov chains from PS1+CMB+BAO.  
We can see again that the best estimate for $H_0t_0$ lies around 1.
The red curve corresponds to a Gaussian PDF with mean =1.001 and standard deviation = 0.03 (see Table \ref{TableCosmoValues}).

\begin{figure}
\begin{center}
\includegraphics[width=7.5cm]{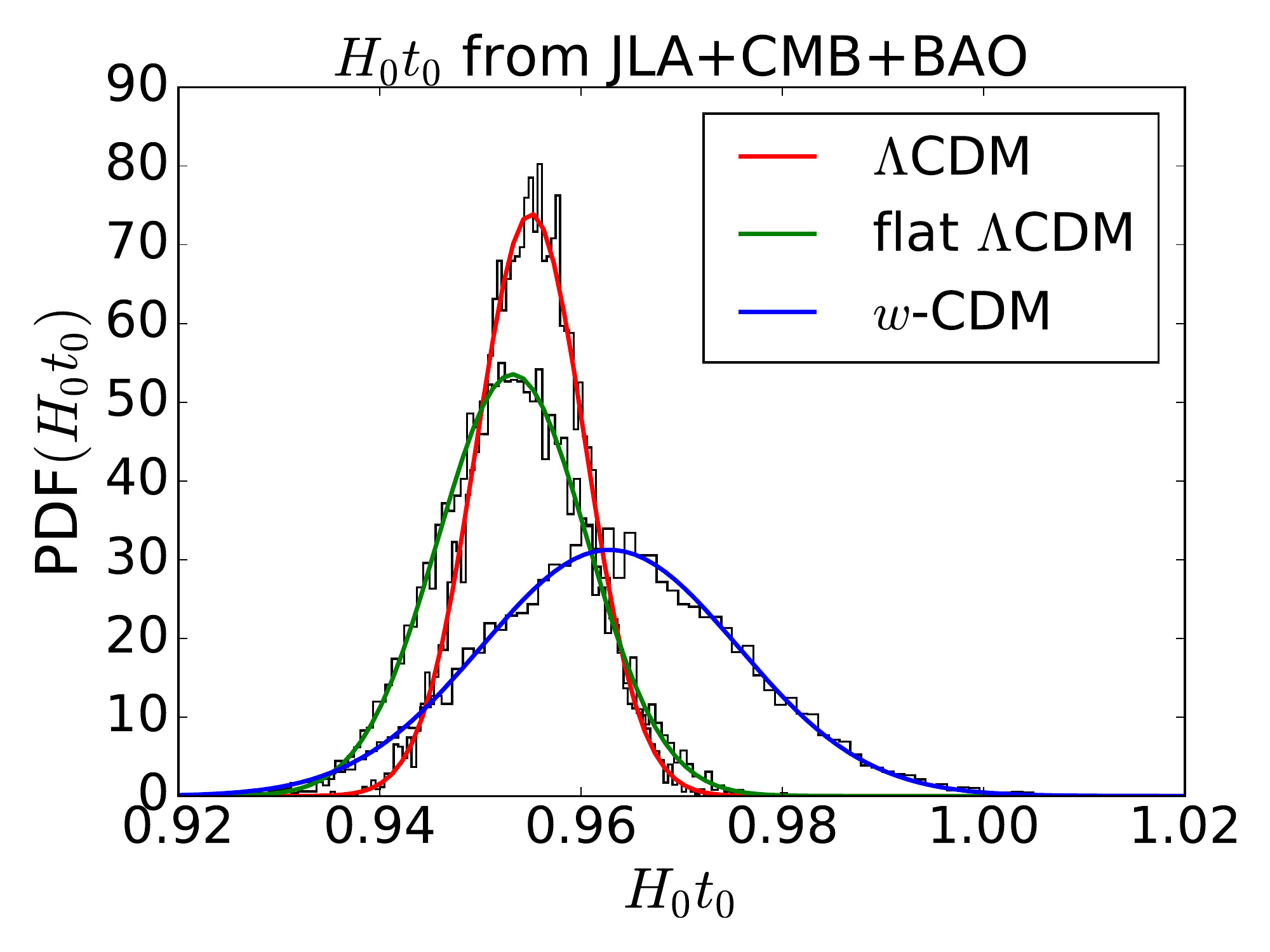}%
\caption{Probability density function (PDF) of $H_0t_0$ computed from $\Lambda$CDM and  extension.
The plots correspond to the use of JLA+CMB+BAO datasets to compute the PDF of $H_0t_0$. See Table \ref{TableCosmoValues} for the mean and standard deviations of  $H_0t_0$ obtained from these PDFs.}
\label{Plot_PDFHoto_JLABAOCMB}
\end{center}
\end{figure}

Figure \ref{Plot_PDFHoto_JLABAOCMB} shows the PDF for $H_0t_0$ from $\Lambda$CDM and its extension, computed from  JLA+CMB+BAO data. 
Table \ref{TableCosmoValues} summarizes the  best estimated values for $H_0t_0$.  We find that for all the combinations of data sets and models, the best estimates for the dimensionless age are close to 1. 
Notice that in this case the constraints on $H_0t_0$ indicate that its value is not \textit{exactly} one, but has a most likely value that is just 5\% lower.

Our point in this paper is that the observations indicate that $H_0t_0$ is very close to one compared with the past and future of the evolution history of the Universe, but not necessary exactly one (it would mean an even worst synchronicity problem).

		\section{Astrophysical constraints on $H_0t_0$}
		\label{SectionAstrophysics}

As we have shown, the information from cosmological constraints, especially from supernovae, places tight constraints on the present value of $H_0t$,  which is  close to 1.
Another way to proceed is to measure $H_0$, not from a cosmological model but from the observed Hubble expansion rate and $t_0$ from astrophysical understanding of the oldest objects in the Universe, obtaining a lower bound to its age. { This is a cosmological-model independent way to determine the dimensionless age of the Universe. }
In the 1980s, this approach showed that there was something missing from the prevailing $\Omega_M =1$ cosmological model, for which $H_0t_0$ was $\frac{2}{3}$, while the measured rate of Hubble expansion and the ages of stars in the oldest globular clusters pointed toward a significantly larger value for $H_0t_0$. 

We denote as $H_{\rm astro}$ and $t_{\rm astro}$ the Hubble constant and age of the Universe measured from astrophysical objects instead of a cosmological model.
The expansion age $t_H  \equiv H_0t_0 / H_{\rm astro}$ can be compared with the bounds on the age  $t_{\rm astro}$.  To put this in units with dimensions, we note that a Hubble constant of $H_{\rm astro} = 73.03 \pm 1.79$ km s$^{-1}$Mpc$^{-1}$ 
measured from the distances to the host galaxies of SN Ia using Cepheids \citep{HzToday_Riess_2016} corresponds to an expansion age of $t_H = 13.39$ Gyr when $H_0t_0 =1$.
The value of $H_{\rm astro}$ estimated by Riess et al.  corresponds to the direct measurement of the Hubble constant in our local Universe by measuring the local velocity field. 

Astrophysical estimates for the ages of objects in the Universe depend on a long train of inference and it is difficult to assess the systematic uncertainties.  
They include  nuclear chronometers, the white dwarf cooling sequence, the shape of Hertzprung-Russell diagrams for globular clusters, and stellar ages inferred for individual stars of exceptionally low metal abundance in the halo of the Milky Way. The ratios of long-lived isotopes like Thorium or Uranium with well-determined half-lives are conceptually appealing chronometers, and those elements have been observed in some stars, with ages originally inferred to be in the range 11-15 Gyr (see for instance \citet{HillEtal2002-AgeStarsFromRadioactivity,FrebelEtal2007-AgeStar}), however uncertainties in the production ratios weaken the constraints on age \citep{Valls2014-StellarAgesAndUniverse}, so it is hard to use this approach to confront the cosmological question.

Simulations suggest that the halo population formed probably 0.2-0.3 Gyr after the Big Bang \citep{RitterEtal2012,SafranekEtal2014} and, since globular clusters belong to that population, they have been used to help bound the cosmic age by fitting theoretical isochrones to the observed color-magnitude diagram.  

Higher precision is claimed for subgiant field stars with extremely low heavy element abundance that are found in the Milky Way halo.  The paucity of heavy elements implies these are old stars that formed before much enrichment of the Galaxy's halo had taken place.  Subgiant stars are on their way from the main sequence to the red giant branch of the color-magnitude diagram, and this is the region where the luminosity of a star of a given color depends most strongly on the age, providing the best leverage for interpreting the observations.  The difficulty is that comparing the theoretical luminosity with the observed magnitude demands an accurate distance and only a few suitable stars are close enough to the Sun to have good parallax measurements.  
For these bright nearby stars, high-resolution spectra can be obtained to help provide accurate chemical abundances and constraints on interstellar reddening that improve the age estimate.  

A recent study using the Fine Guidance Sensors on the Hubble Space Telescope to determine the parallax gives ages and the uncertainty due to the parallax for 3 of these stars:  HD 84937, HD 132475, and HD 140238  of $12.08 \pm 0.14$ Gyr, $12.56 \pm 0.46$ Gyr and $14.27 \pm 0.38$ Gyr respectively \citep{VandenBergEtal2014}.  They also estimate that the uncertainty from other sources has an effect of about $\pm 0.8$ Gyr. If they have correctly gauged the errors, this is beginning to be interesting.  Since we can expect the GAIA space mission to produce high-precision parallax measurements for these and other halo subgiants, a confrontation between astrophysical ages and cosmological ones may evolve into a useful tool for cosmology.  Just as adopting $H_0$ as a prior in a cosmological analysis can have a substantial effect on the overall inference (see for instance \citet{CMB_Planck_Cosmology_2015}) establishing a well-determined age for the universe could be used to constrain the best model.

Cosmology is based on gravity as formulated in general relativity:  but the ages of stars depend on the strong, weak, and electromagnetic forces acting to produce nuclear burning following the laws of quantum mechanics. Because general relativity  and quantum mechanics are separate theories, it is quite amazing that they converge on the same age for the Universe.

We use the age of the stars HD 84937, HD 132475, and HD 140238  to compute the corresponding lower limit on the dimensionless age of the Universe from astrophysical objects and find out how close are those values of one, and also to compare with the cosmological predictions shown on table \ref{TableCosmoValues}.

\begin{figure}
\begin{center}
\includegraphics[width=7.5cm]{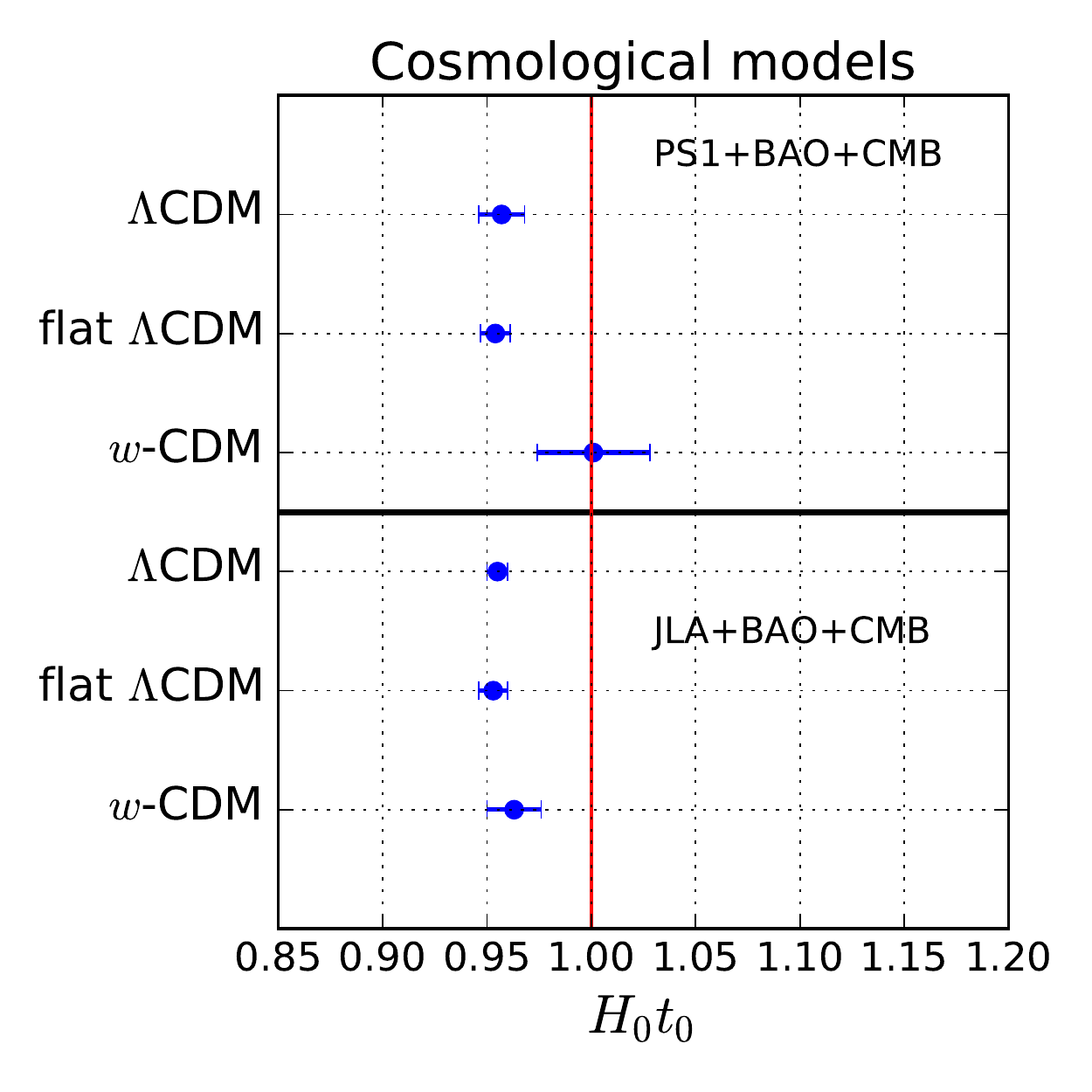}
\includegraphics[width=7.5cm]{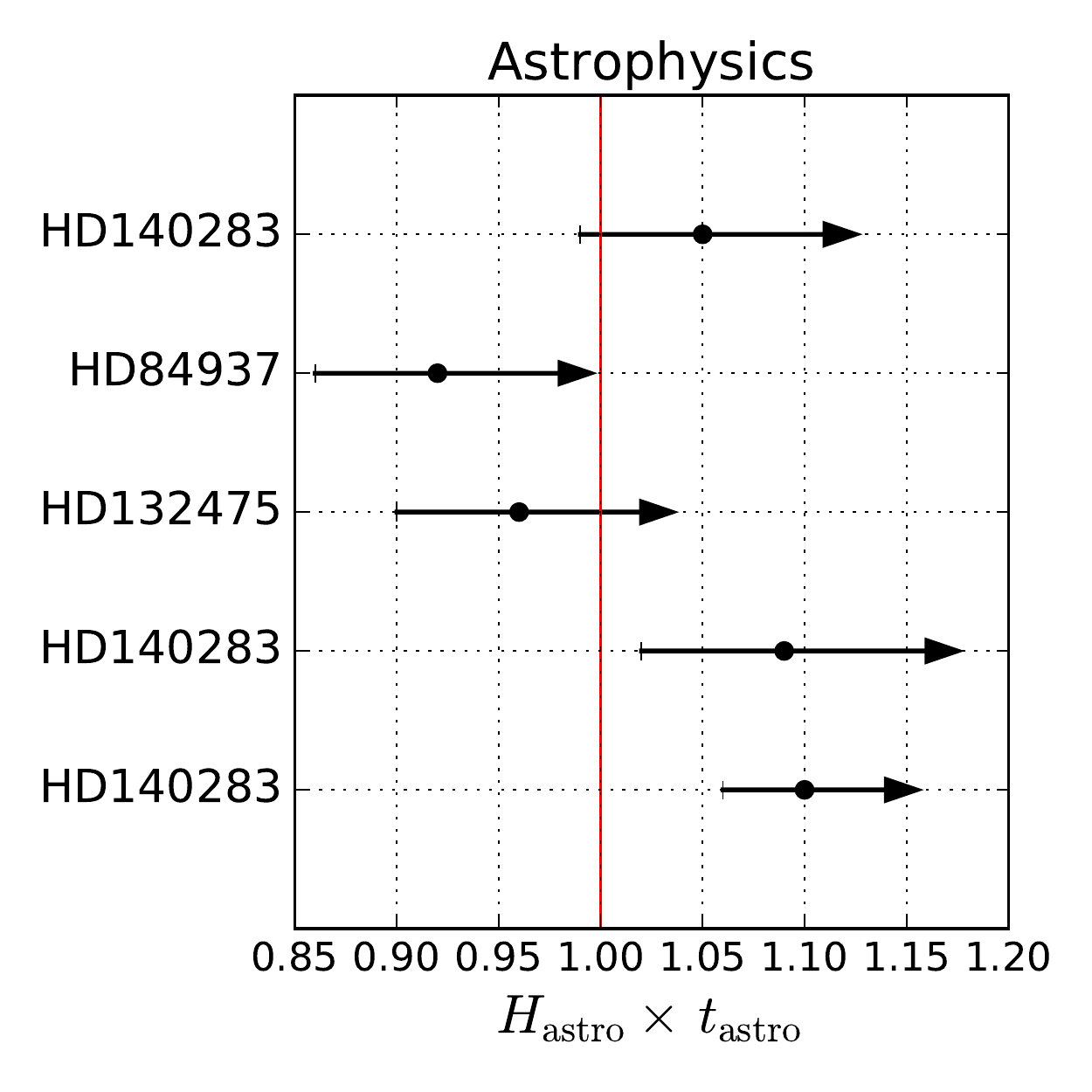}
\caption{ { Comparison of constraints on the dimensionless age from cosmological models and astrophysical objects in the upper and lower panels respectively. 
The blue intervals in the upper panel correspond to the $1\sigma$ estimation of  $H_0t_0$ shown in column 5 of table \ref{TableCosmoValues}. And the black arrows in the lower panel correspond to the $1\sigma$  intervals shown in column 4 of table \ref{TableAgeObserved},  they are the estimation of the age of the Universe from astrophysical objects. We use arrows instead of bars to emphasize that the upper constraint of the age is unbounded.}
}
\label{PlotAgeGyrLCDMAstrophysics}
\end{center}
\end{figure}

Table \ref{TableAgeObserved} shows the results assuming $H_{\rm astro} = 73.03 \pm 1.79$ km s$^{-1}$Mpc$^{-1}$.
We find that in all cases the best estimated values for $H_{\rm  astro} \times t_{\rm  astro}$ are around one but with more dispersion and larger uncertainties than the cosmological model estimations.
{ Figure \ref{PlotAgeGyrLCDMAstrophysics} illustrates the cosmological and astrophysical estimates of the dimensionless age.}

\begin{table*}
\centering
\begin{tabular}{ l l c  c  }
\multicolumn{4}{c}{\textbf{Ages of oldest stars}} \\
\hline \hline
Star & $t_{\rm astro}$ (Gyr) & Ref.  & $H_{\rm astro} \times t_{\rm astro}$   \\ 
 \hline
 \hline
  HD 140283 & $ 14.00 \pm 0.7$ & \citep{CreeveyEtal2015}  & $1.05 \pm 0.06$  \\ 

 HD 84937  & $ 12.38 \pm 0.8$ & \citep{VandenBergEtal2014} & $0.92 \pm 0.06$ \\ 

 HD 132475 &  $ 12.86 \pm 0.8$ & \citep{VandenBergEtal2014}  & $0.96 \pm 0.06$   \\ 

 HD 140283 & $ 14.57 \pm 0.8$ & \citep{VandenBergEtal2014}  & $1.09 \pm 0.07$   \\ 

 HD 140283 & $14.76 \pm 0.3$ & \citep{AgeUniv-BondEtal-2013}  &  $1.10 \pm 0.04$   \\
\hline
\end{tabular}
\caption{  Constraints on the age of the Universe from astrophysical objects. The first and second columns show the name of some of the best known oldest stars and the estimation of the age of the Universe based on these stars. We have added 0.3 Gyr to the reported age of each star to account for the time elapsed between the Big Bang to the formation of the first stars in the Universe \citep{RitterEtal2012,SafranekEtal2014}.
The 4th column shows the dimensionless age $H_{\rm astro} \times t_{\rm astro}$ assuming the value of $H_{\rm astro}  = 73.03 \pm 1.79$ km s$^{-1}$Mpc$^{-1}$. Uncertainties are the 68.3\% confidence intervals.
}
\label{TableAgeObserved}
\end{table*}


		\section{Conclusions}\label{SectionConclusions}	

When the dimensionless age of the Universe, $H_0t_0$,  is calculated using the $\Lambda$CDM model and the corresponding best estimated values for its cosmological parameters computed from the observations, it turns out that $H_0t_0$ is close to 1 at the \textit{present} time. 
{ It means that the current age of the Universe is very close to the Hubble time $1/H_0$. }

In other words, along the evolution of the Universe the dimensionless age had and will have values in the very wide range $0<H_0 t(a)< \infty$,
however,  it appears that we are living \textit{today} in a very ``special'' time when $H_0 t_0 \sim 1$ according to the $\Lambda$CDM model and the current cosmological observations, even when the possible range of the \textit{present} values of the dimensionless is also as large as $2/3<H_0 t_0< \infty$.
{We call this the ``{synchronicity} problem'' of the age of the Universe.}

{This synchronicity problem means that the decelerated expansion epoch of the early Universe is compensated with the late time accelerated epoch such that the total expansion today is similar to the case if the Universe were expanding at the same expansion rate since the Big Bang.
Even more, this ``coincidence'' only happens \textit{today}.  For any other time in the past or future, the age of the Universe will not be so nearly the Hubble time as it is today. }

This problem is related to the well-known coincidence problem, but it is not identical.  There can be combinations of $\Omega_M$ and $\Omega_{\Lambda}$ for which the coincidence problem holds, but for which $H_0 t_0$ is not close to one.
For instance, consider the hypothetical case  where the matter-energy content of the Universe were  $\Omega_{M} = 0.15$ and  $\Omega_{\Lambda} = 1$. Here $\Omega_M \sim \Omega_{\Lambda}$, but $H_0 t_0 = 1.3$ is not so close to 1.0  as in the case of the concordance values of $(\Omega_{M}, \Omega_{\Lambda}, w)$.
{ From a perspective independent of cosmological models; if the value of the Hubble constant were very different from $\sim 70$ km s$^{-1}$ Mpc$^{-1}$  then the current age of the Universe would be very different from the Hubble time. }

We considered the latest type Ia SN samples, Pan-STARRS  and JLA, combined with the anisotropies of the CMB and the baryonic acoustic oscillations, in order to infer the  value of $H_0t_0$ from $\Lambda$CDM and extensions. We found that in all the cosmological models and combinations of data sets $H_0t_0 \sim 1$. 

{ We investigate the predictions about the dimensionless age from astrophysical objects. These inferences are independent of cosmological models.  We found again that $H_{\rm astro} \times t_{\rm astro} \sim 1$ in all the cases, though the  constraints are less tight than the $\Lambda$CDM constraints.}

The reason why $H_0 t_0 \sim 1$ remains unclear: it might be a meaningless coincidence or it might be a clue to something missing from our present understanding.

Finally, as \citet{HzToday_Riess_2016} suggest, based on the difference between the value of $H_0$ they measure and the value that makes the distance to the last scattering surface fit the CMB power spectrum there could be something missing from the $\Lambda$CDM cosmological model. An extra relativistic particle like a sterile neutrino or a time-variable dark energy with $w$ not equal to 1 are possibilities that could modify the expansion history of the Universe.  When we have more secure ages for the stars, we will be able to test these ideas against the evidence in another way using the age of the universe.


\section{Acknowledgments}
The authors thank to Adam Riess, Daniel Eisenstein, Eric Linder, Kaisey Mandel, Andrew Friedman and Raul Jimenez for very useful discussions and suggestions to improve this work. 
Supernova cosmology at the Harvard College Observatory is supported by National Science Foundation grants  AST-156854 and AST-1211196. 
A.A. acknowledges also, the Instituto Avanzado de Cosmolog\'ia of Mexico, and  the Mexico-Harvard Fellowship  sponsored by Fundaci\'on M\'exico en Harvard and CONACyT.
We acknowledge the use of \textit{emcee}  \citep{MCMCSoftware-emcee}. 


	\appendix
	
	\section{Determination of $H_0t_0$ from the MCMC Markov chains}
	\label{Sec_MCMCMarkovChains}
	
\begin{figure}
\begin{center}
\hfill
\includegraphics[width=8.2cm]{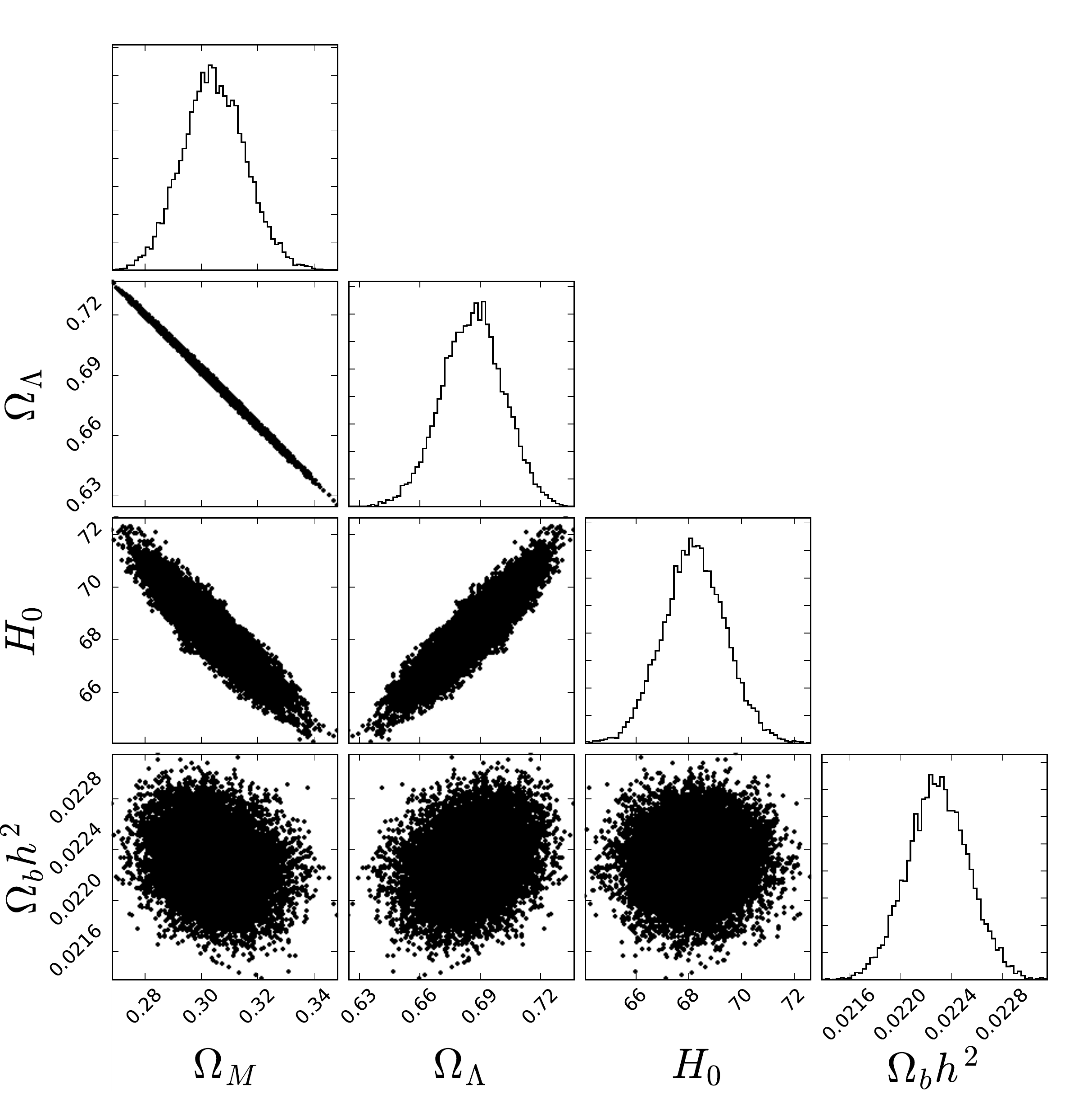}%
\hfill
\includegraphics[width=8.2cm]{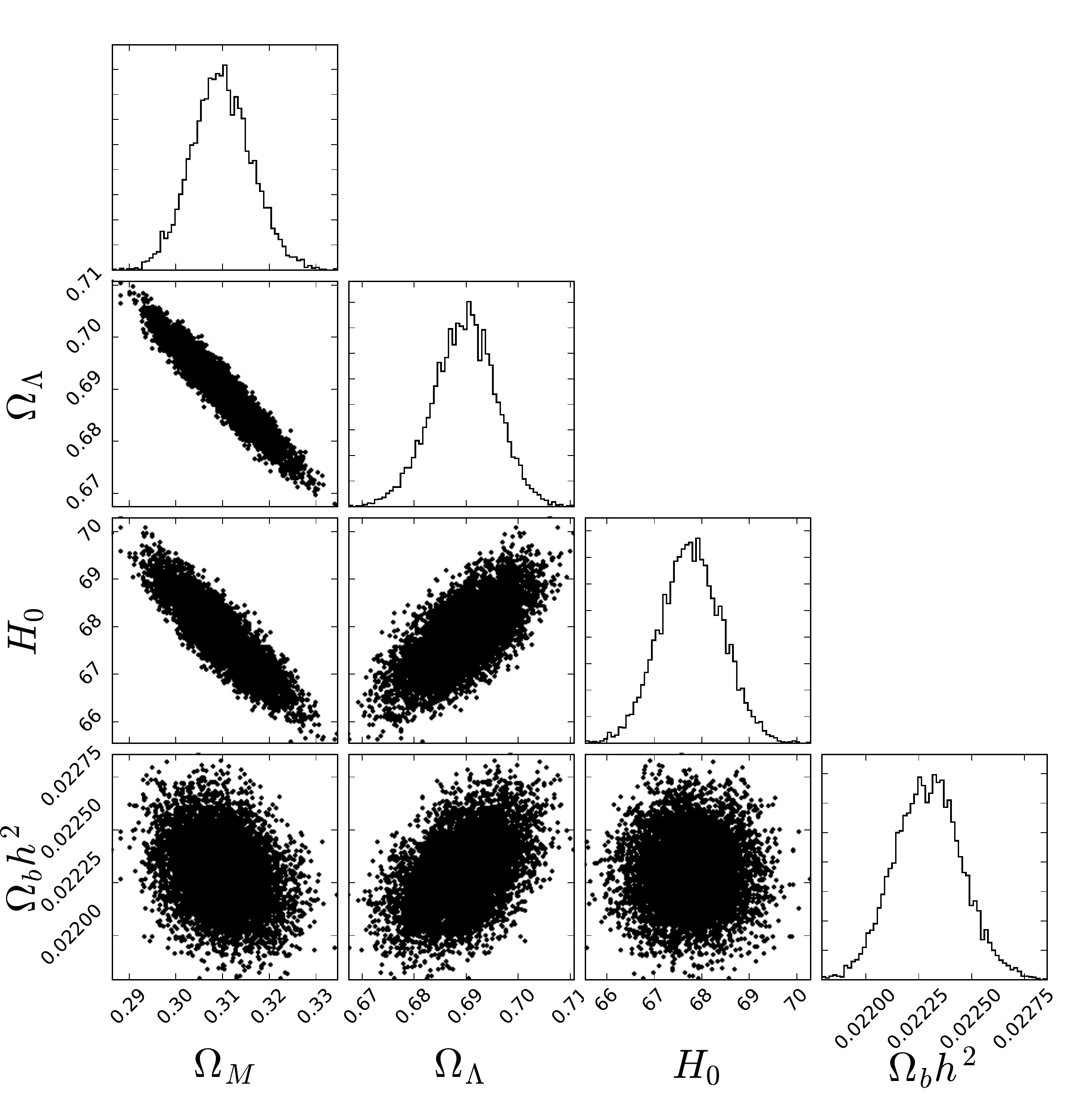}
\hspace*{\fill}
\caption{  MCMC Markov chains used to compute the PDF of $H_0t_0$ from $\Lambda$CDM.
Left and right panels shows the Markov chains computed using PS1+CMB+BAO  and JLA+CMB+BAO datasets respectively. For PS1+CMB+BAO we computed the Markov chains and for JLA+CMB+BAO we used those already computed by \citet{CMB_Planck_Cosmology_2015} and available at 
\href{http://pla.esac.esa.int/pla/\#cosmology}{http://pla.esac.esa.int/pla/\#cosmology}. 
}
\label{PlotMCMCMarkovChains}
\end{center}
\end{figure}

We determine the PDF for $H_0t_0$ by using the MCMC Markov chains computed to estimate the cosmological parameters combined with Eqs. (\ref{EqIntegralForHoto}) and (\ref{EqIntegralForHotoCPL}). 
Figure \ref{PlotMCMCMarkovChains} shows the MCMC Markov chains used to  constrain $\Lambda$CDM from the PS1+CMB+BAO (left panel) and JLA+CMB+BAO (right) datasets.
So for instance, for every step in the chain shown in the left panel of Fig. \ref{PlotMCMCMarkovChains} we insert the values of the cosmological parameters of that step in Eq. (\ref{EqIntegralForHoto}) obtaining a single value of $H_0t_0$. Then repeating this procedure for the full chain we obtain the histogram for $H_0t_0$  shown in black in Fig. \ref{FigureOmOLPDF}.
A similar procedure was used to determine the PDF of  $H_0t_0$ from the flat $\Lambda$CDM and $w$-CDM models.
With this procedure the full covariance among the cosmological parameters is taken into account to determine $H_0t_0$.
We find that the PDFs of $H_0t_0$ have a Gaussian profile. Column 5 of Table \ref{TableCosmoValues} shows the mean and the standard deviation of each PDF of  $H_0t_0$.

\end{document}